\documentclass[twocolumn]{aastex6}
\usepackage{graphicx}
\usepackage{cellspace}
\usepackage{enumitem}
\newcounter{descriptcount}

\newcommand\cincludegraphics[2][]{\raisebox{-0.3\height}{\includegraphics[#1]{#2}}}
\newcommand{\inline}[1]{\includegraphics[height=11pt]{#1}}
\newcommand{\capline}[1]{\includegraphics[height=8pt]{#1}}

\newcommand{\Fig}[1]{Figure~\ref{#1}}
\newcommand{\Tab}[1]{Table~\ref{#1}}
\newcommand{\Sect}[1]{\S~\ref{#1}}

\newcommand{\sex}{\texttt{SExtractor}}

\begin{document}

\title{A Graphical-User Interface for Editing Segmentation Images}
\shorttitle{Segmentation Editor}
\shortauthors{Ryan Jr.}

\author{R.~E.~Ryan Jr.}
\affil{Space Telescope Science Institute\\
3700 San Martin Ave.\\
Baltimore, MD 21210, USA}

\begin{abstract}
  I present a graphical-user interface for performing several image
  operations on segmentation maps.  The package is written entirely in
  IDL, and is provided as source code (for those who may want to
  develop, link to existing packages, or reappropriate the code base)
  and eventually as a stand-alone, run-time executable upon request
  (for those without an IDL license).  The software facilitates a
  number of operations, which are generally tedious without a
  graphical interface, such as deleting, merging, ungrouping, and
  drawing regions; erasing and painting individual pixels; and
  compression, dilation, and erosion of a segmentation image.  The
  segmentation image is displayed with random RGB triplets to ensure
  adjacent regions are readily discernible, whereas the direct image
  is shown as an inverted greyscale with controls for brightness
  range, bias, and contrast with several scaling functions (as similar
  to \texttt{ds9}).  The opacity between the segmentation and direct
  image is tunable, which gives full control to the image display.

\end{abstract}

\keywords{methods: data analysis}

\section{Introduction}\label{sec:intro}

Image segmentation refers to the process of uniquely identifying a
given pixel or set of pixels in a digital image as a member of a
region.  In astronomy, this notion led to the development of
segmentation images (or maps) that define astrophysical sources in
pixelated images.  Such images are often used used to define apertures
through which various measurements are made (for example brightness or
center-of-mass).  Therefore, it is imperative to ensure that the
segmentation maps are free of pathological defects and false positives
are judiciously removed.

One  common tool  for  algorithmically creating  segmentation maps  is
\texttt{Source Extractor}  \citep[\sex;][]{sex}, which  has a  host of
parameters  that govern  the properties  of the  deblending of  nearby
sources. Although \sex\  is extremely efficient, it  can often segment
an image in  a way far different  than a human might  expect or desire
(notably regions near extended sources  or bright point sources).  But
also,  it  is  challenging  to force  a  segmentation  region  without
inadvertently creating  many false  sources.  To remedy  these issues,
several schemes have proposed to remove extraneous sources \citep[such
  as  hot/cold running;][]{bard12}.   Here, I  propose an  alternative
paradigm: a  graphical-user interface  (GUI) that  allows the  user to
directly interact with the  segmentation image pixels.  Although there
are  distinct  advantages  with   an  algorithmic  approach  (such  as
repeatability), a  properly cleaned segmentation map  avoids erroneous
sources  corrupting the  aggregate  properties of  the sample  and/or
biasing  measurements  of other  sources  in  the field.   With  these
tradeoffs in  mind, I  expect the  ideal approach is  to begin  with a
segmentation  map  derived by  some  algorithmic  means then  sensibly
polishing the map for clear mistakes, but \textit{primum non nocere}.

This document is organized as follows: I introduce the package,
including installation and functionality in \Sect{sec:editor}, I
present several examples in \Sect{sec:examples}, I describe the User
preferences in \Sect{sec:pref}, I present additional improvements
and limitations of the current implementation in \Sect{sec:add}, and
close with a few closing remarks and waivers in \Sect{sec:coda}.

\section{Segmentation Editor}\label{sec:editor}

\subsection{Installation, Dependencies, and Calling}
The         GUI         is         entirely         written         in
IDL\footnote{\href{http://www.harrisgeospatial.com/SoftwareTechnology/IDL.aspx}{http://www.harrisgeospatial.com/SoftwareTechnology/IDL.aspx}},
and  is  installed  by  simply  placing  the  directory  tree  in  the
\texttt{IDL\_PATH}  of  the user.   The  code  requires the  astronomy
library    \citep[\texttt{AstroLib};][]{astrolib},   which    can   be
similarly  installed. The  main  component  of \texttt{SegEditor}  is
written as a class that is instantiated from the IDL command line as:
\begin{verbatim}
IDL> obj=obj_new('segeditor',SEGFILE,IMGFILE)
\end{verbatim}
\vspace*{1ex}
\noindent  where \texttt{SEGFILE}  and \texttt{IMGFILE}  are variables
referring  to the  full  path  of the  segmentation  and direct  image,
respectively.   I  also  provide a  procedural-oriented  wrapper  that
handles the object-oriented aspects  internally.  Since older versions
of IDL  do not  explicitly destroy  objects, it  may be  imperative to
explicitly  destroy the  resultant object  (\texttt{obj} in  the above
example).  However,  if the  GUI is properly  closed, then  the object
reference will also be  destroyed.  Therefore if \texttt{SegEditor} is
called from an  external code-base, it may be necessary  to verify the
object still exists (such as with \texttt{obj\_valid()}).

\subsection{Basic Functionality}

\begin{figure}
%  \plotone{segeditor.ps}
  \includegraphics[width=3.3in]{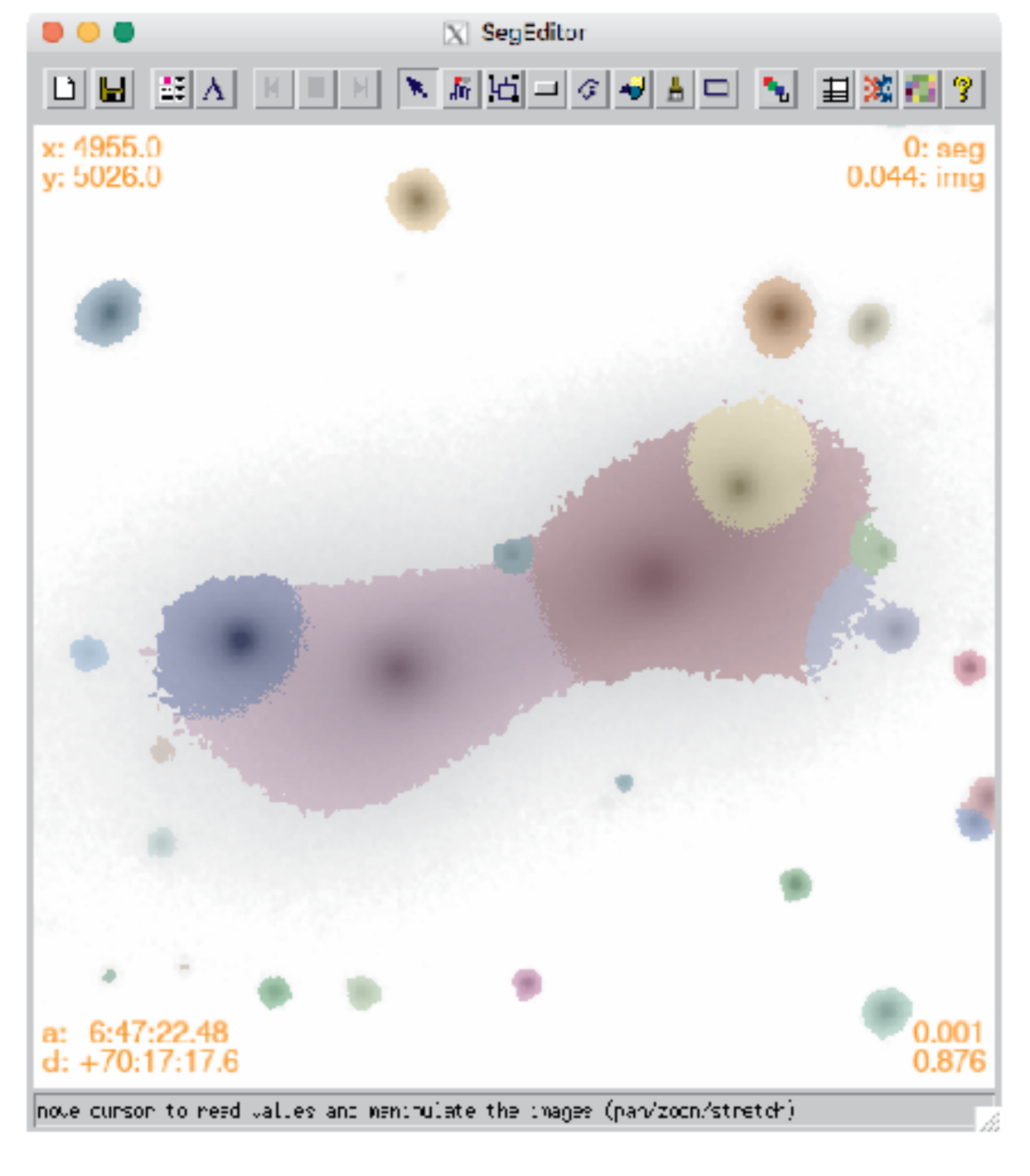}
  \caption{Basic state of the \texttt{SegEditor}  GUI.  At the top row
    is the  primary toolbar in the  default state.  In the  middle, is
    the  main   image  display   with  the   segmentation  map   as  a
    semi-transparent three-color and the  direct image as a gray-scale
    image,  respectively.   The opacity  of  the  segmentation map  is
    controlled    by    the    right-mouse   button    described    in
    \Tab{tab:buttons}.  In  the four  corners of the  graphics window,
    the $(x,y)$ positions, image  pixel values, min/max display values
    for  the direct  image, and  the RA/Dec  positions (from  top left
    going  clockwise).   Finally,  at  the  very  bottom  is  a  brief
    instruction of the current left-mouse state.
    \label{fig:segeditor}}
\end{figure}

The initial state of the GUI is shown in \Fig{fig:segeditor} with data
from \citet{clash}.   There are several  quick keys to  execute common
functions,   often  for   the   left  side   of   the  keyboard   (see
\Tab{tab:keyboard}).   However, the  main controls  are driven  by the
state of the mouse, which is dictated by toolbar along the top.  These
buttons are briefly described in \Tab{tab:buttons}.

\begin{deluxetable}{ll}
  \tablecaption{Keyboard Controls\label{tab:keyboard}}
  \tablehead{\colhead{Key} & \colhead{Description}}
  \startdata
  SPACE & Step the left-mouse state\\
  d & Redisplay the image to the screen\\
  u & Undo the last region manipulation, same as \cincludegraphics[height=11pt]{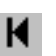}\\
  r & Redo the last region manipulation, same as \cincludegraphics[height=11pt]{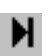}\\
  q & Quit the GUI\\
  Q & Force quit the GUI\\
  p & Print the current image to postscript\\
  $\blacktriangle$ & Move cursor up one pixel\\
  $\blacktriangledown$ & Move cursor down one pixel\\
  $\blacktriangleleft$ & Move cursor left one pixel\\
  $\blacktriangleright$ & Move cursor right one pixel\\
  \enddata
\end{deluxetable}

%The main controls are driven by the toolbar buttons, which are
%described in \Tab{tab:buttons}.  There are also several quick keys
%that facilitate actions, often for the left side of the keyboard (see
%\Tab{tab:letters}).

\section{Example Actions}\label{sec:examples}

Although it may be readily straightforward to manipulate the GUI, it may
be useful to describe a few actions to begin.

\begin{figure*}
  \begin{center}
    \includegraphics[width=1.62in]{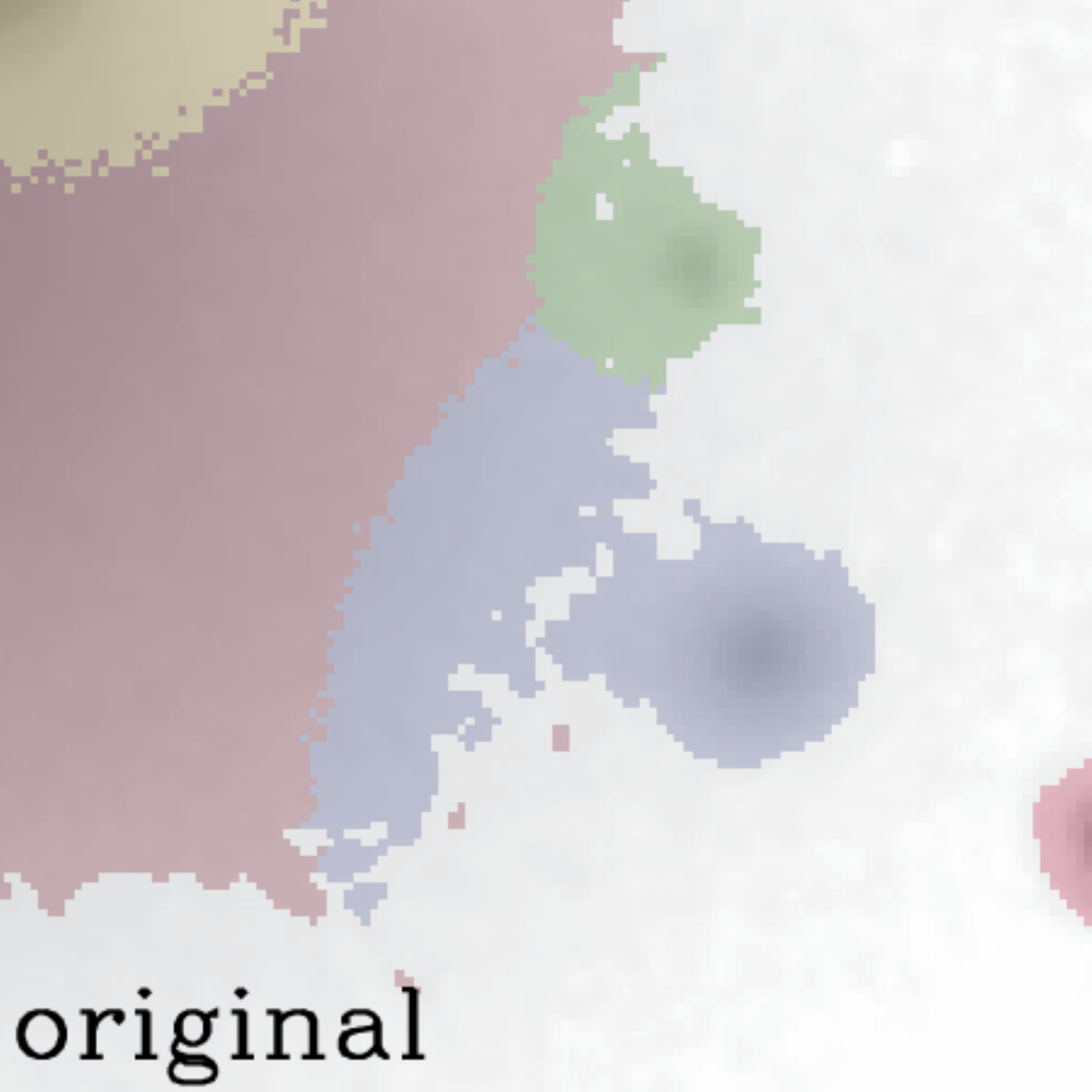}
    \includegraphics[width=1.62in]{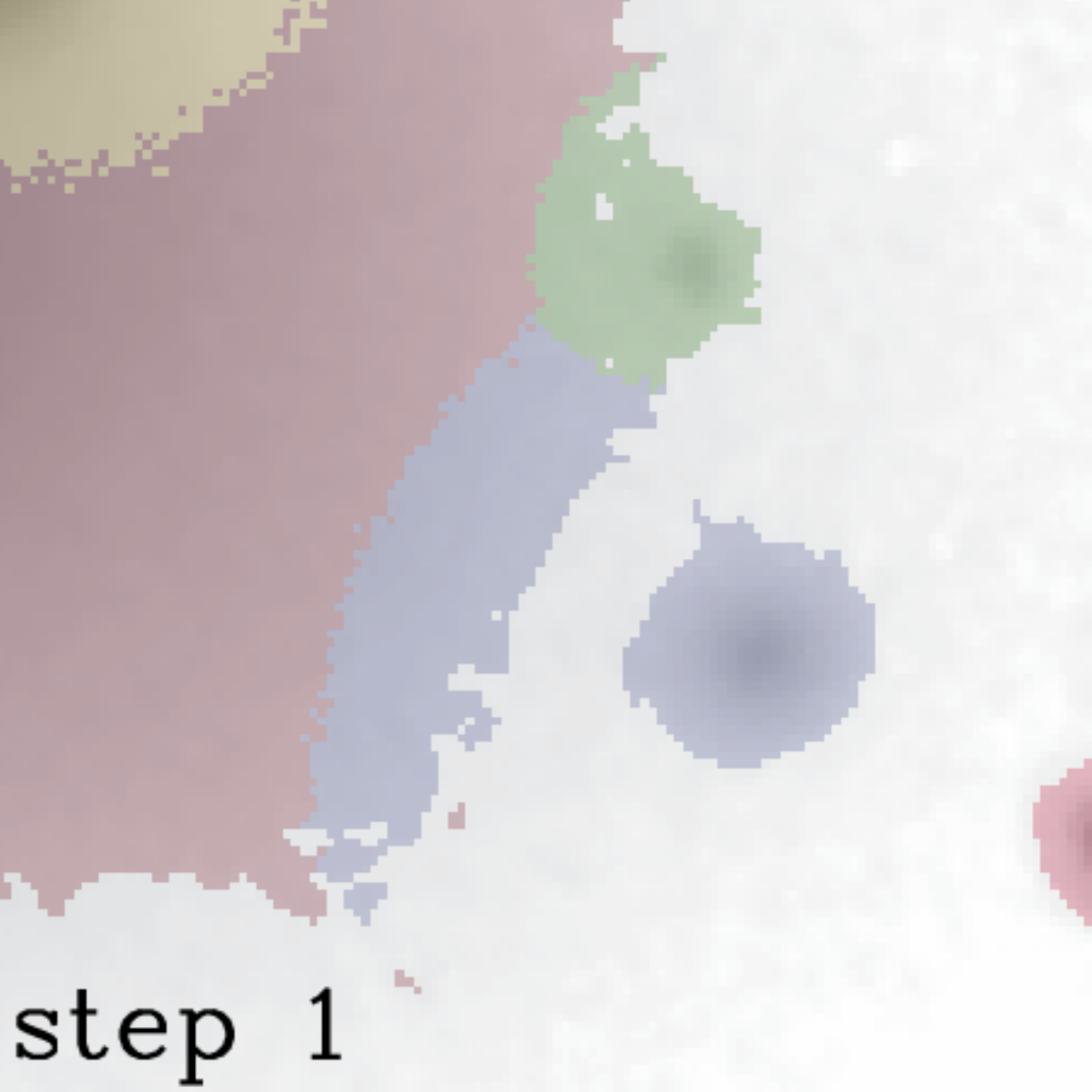}
    \includegraphics[width=1.62in]{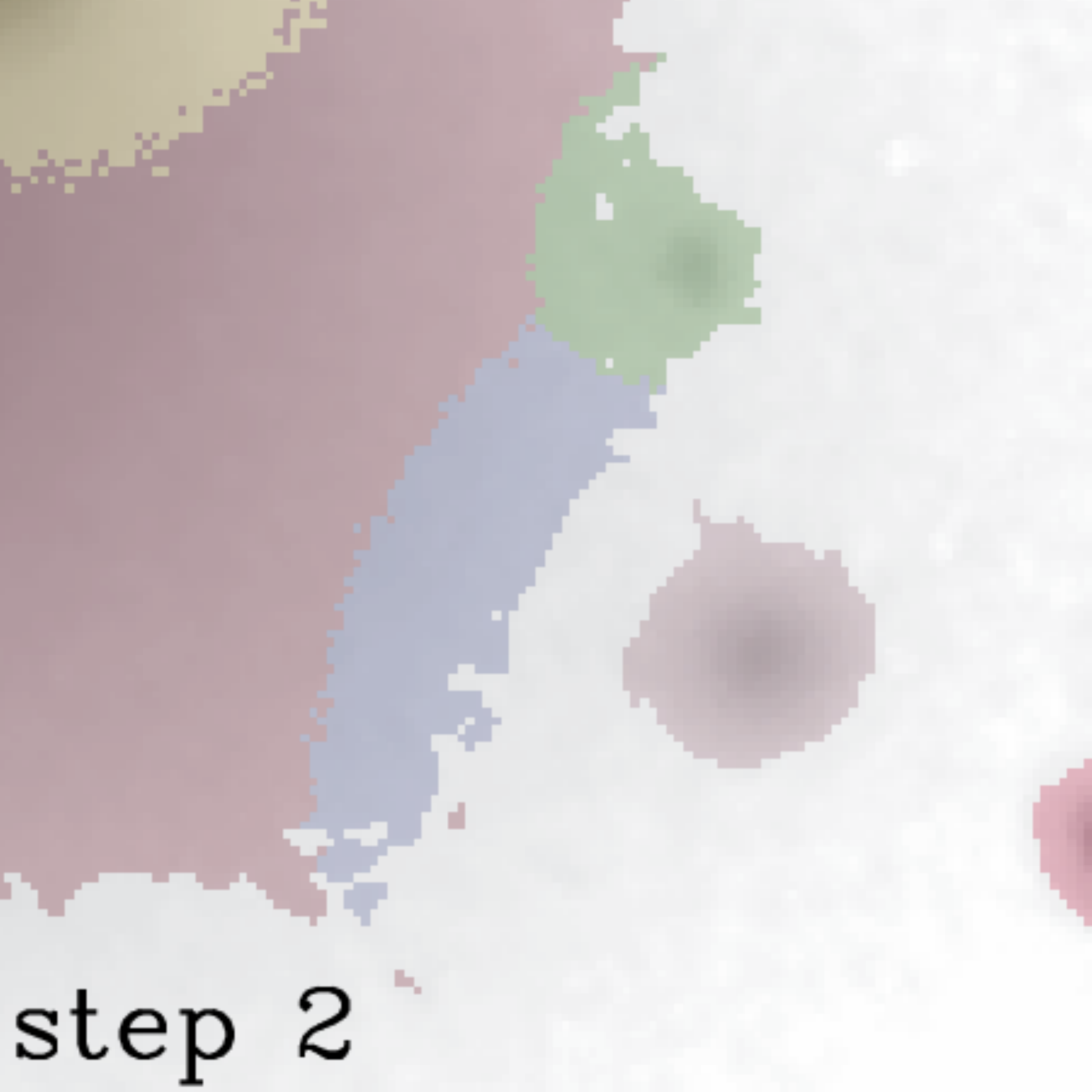}
    \includegraphics[width=1.62in]{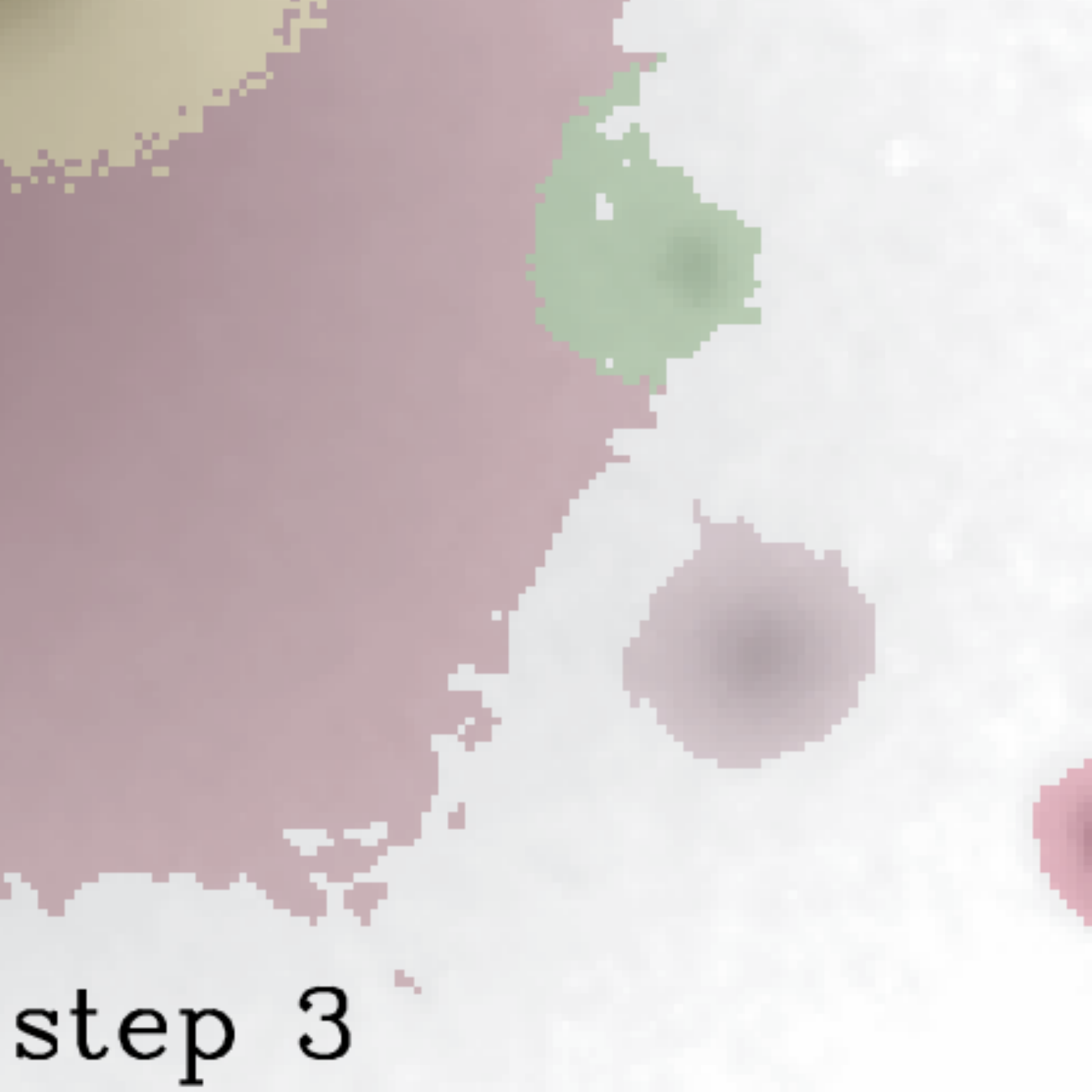}
    \caption{Example to  separate two regions  and regroup.  On  the far
      left, I show the original image (a zoom from \Fig{fig:segeditor}).
      The first step is to  use the eraser tool \capline{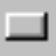}
      to separate the light blue region from the main pink region.  Next
      is  to   {\it  ungroup}  the   separated  blue  region   with  the
      disassociate   tool   \capline{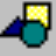}.    Finally,   the
      remaining light  blue region can  be grouped with the  larger pink
      region                with               merge                tool
      \capline{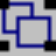}.\label{fig:example}}
  \end{center}
\end{figure*}

In \Fig{fig:example},  I show an example  in using \texttt{SegEditor}.
Here the  light-blue object has been  associated as a small  object to
the  lower right,  but \texttt{SExtractor}  identified several  pixels
that probably should belong to the main galaxy to the upper left (pink
source).  To reassign those pixels to  the main galaxy, one can simple
erase  the collection  of pixels  that  connect the  two regions  with
\inline{button.pdf}.  Once the smaller object is isolated from the
erroneous pixels to the upper left,  then it can be ungrouped from the
pair with  \inline{drawing.pdf} and finally  two bits of  the main
galaxy can be merged with \inline{group.pdf}.

\begin{figure}
  \begin{center}
    \includegraphics[width=1.62in]{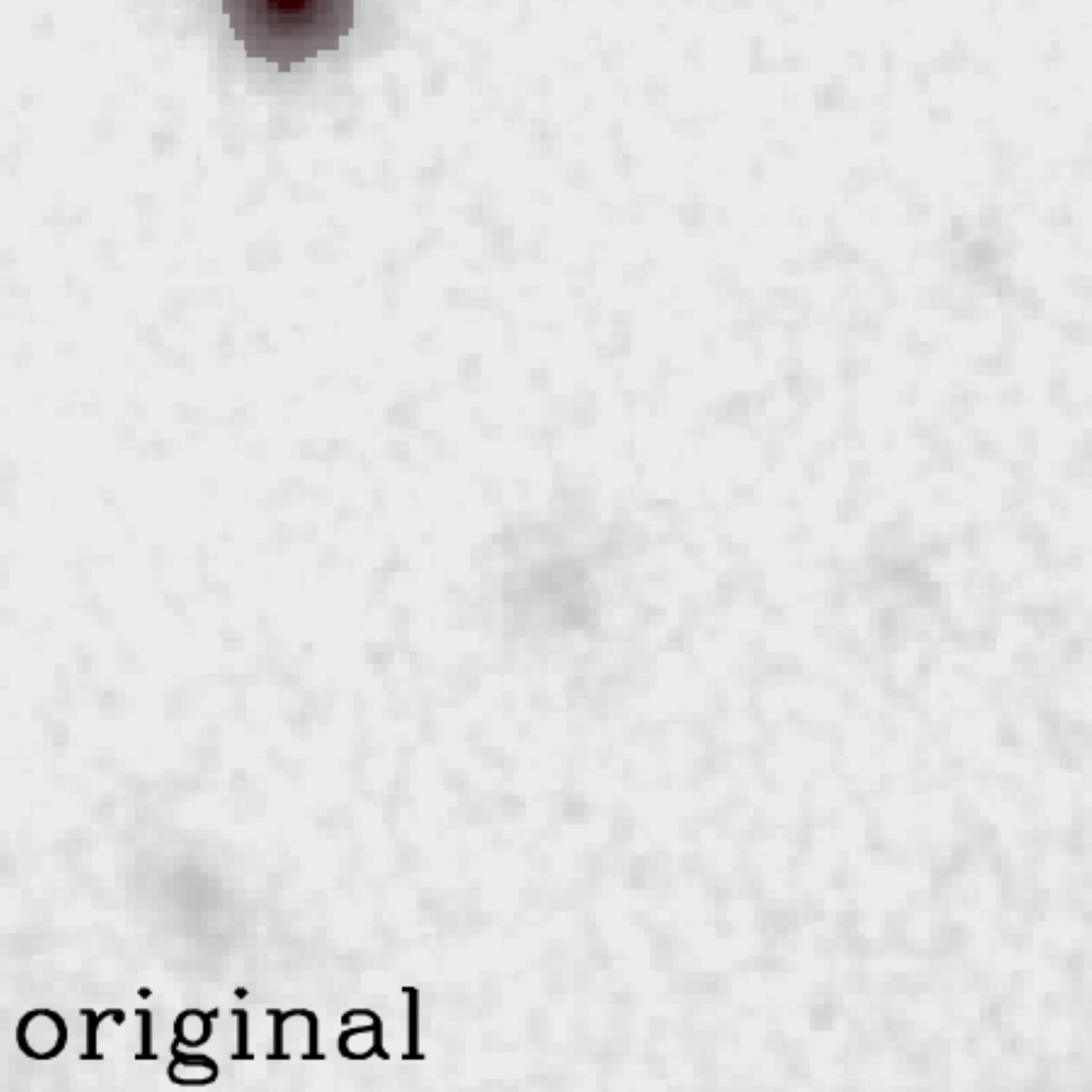}
    \includegraphics[width=1.62in]{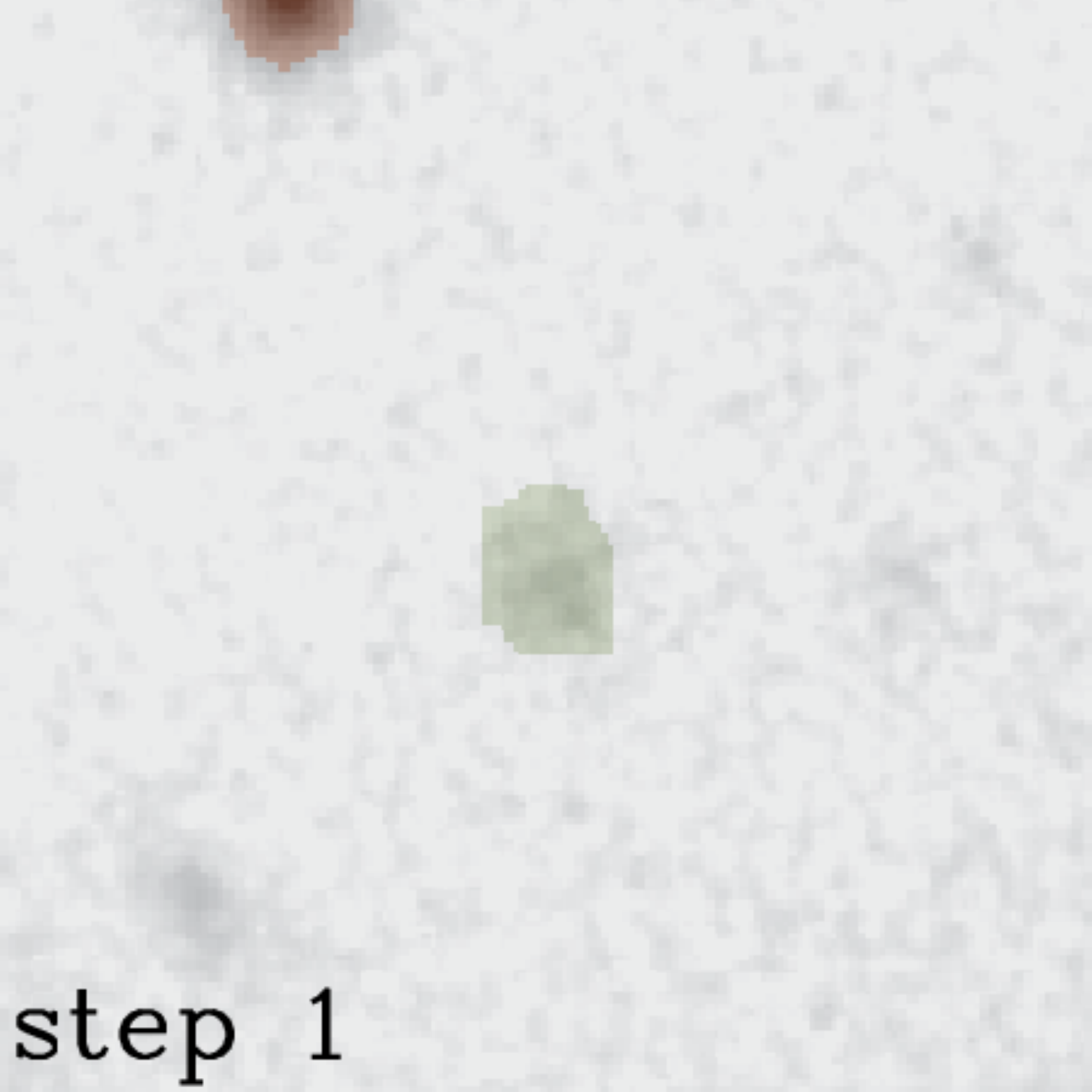}
    \caption{Example  to  draw  a  region.  The  first  image  is  the
      original, again a zoom  from \Fig{fig:segeditor}.  But since the
      source in the center was unidentified by \texttt{SExtractor}, it
      is   necessary  to   create  a   region  with   the  draw   tool
      \capline{segpoly.pdf}.\label{fig:draw}}
  \end{center}
\end{figure}

But it  is perhaps still  the case that one  may wish to  increase the
outer edge  of the region of  a source, such  as that as seen  for the
pink galaxy.  To execute such an operation, one selects the paint tool
\inline{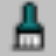} then will  click and drag to  paint the pixels.
However it is important to begin  the click/drag motion on the region
to  be expanded  (the paint  tool  does not  paint a  new region  from
scratch).  To draw  a new region, where one was  not there before, the
draw  tool  \inline{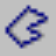}  is  used  (see  \Fig{fig:draw}).
Here, one clicks  and drags to encompass an set  of pixels (the region
is automatically  closed), which will  be given a  unique segmentation
value (the current maximum segmentation value plus one).

Sometimes it is useful to identify a particular known source or scan
the segmentation for pathological issues (such as the smallest/largest
sources).  Such operations are easily achievable from the tabulation
view, which is opened with \inline{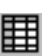}.  With this interface,
one can click on a column heading to sort the table by this column,
and once the data is sorted a second click will sort in reverse order.
Also, by clicking on a row header, the primary GUI will pan to the
object of interest. However it is important to note, the tabulation
GUI will not update as operations are performed in the main window.
Therefore it is imperative to refresh the table with
\inline{dm.pdf}.

\begin{figure}
  \begin{center}
    \includegraphics[width=3.25in]{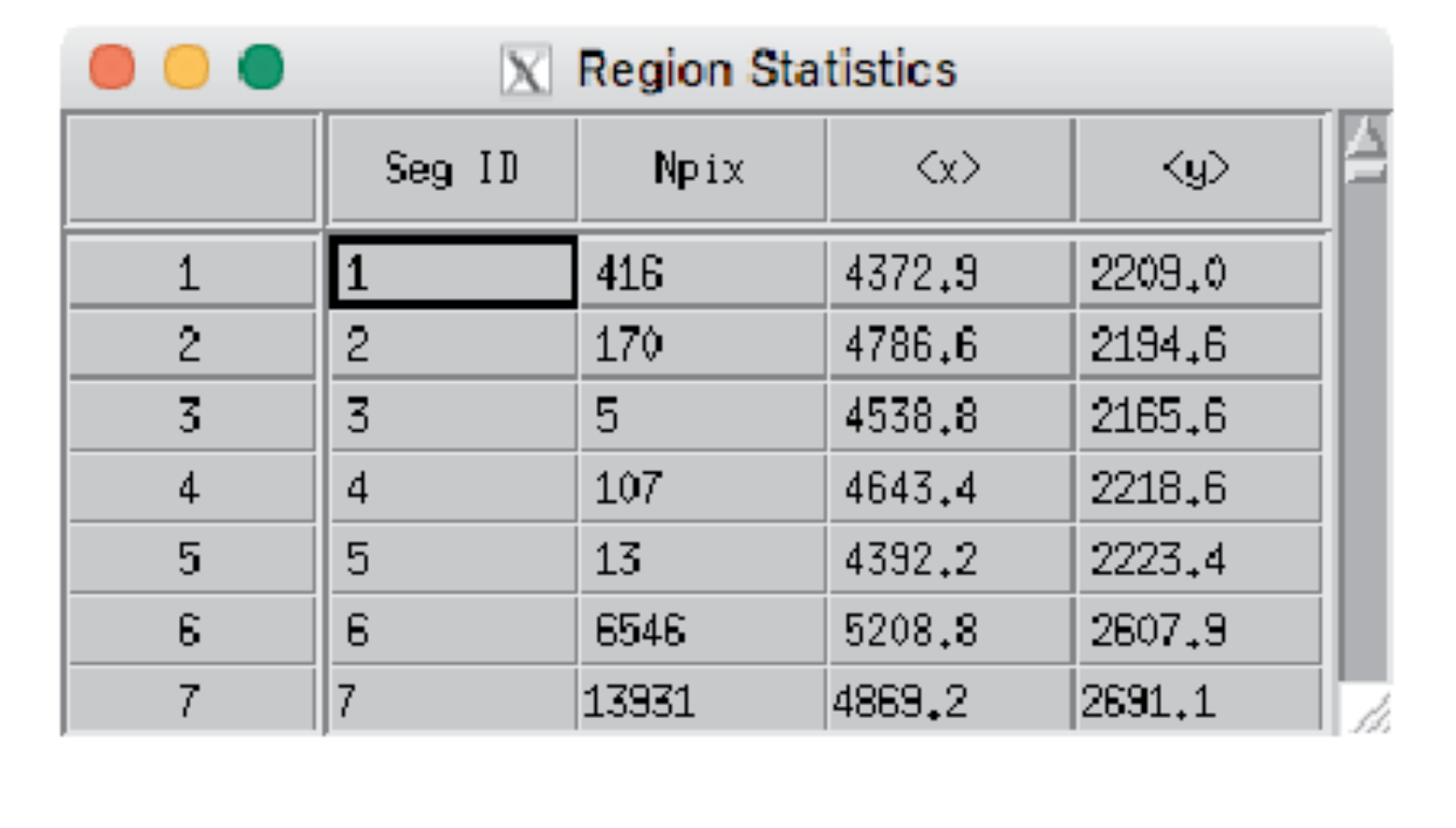}
    \caption{Table view.   The columns can be  sorted either ascending
      or descending by clicking on the column headers.  By clicking on
      a row header, the main  GUI (\Fig{fig:segeditor}) will center on
      that source.\label{fig:table}}
  \end{center}
\end{figure}

\section{Preferences} \label{sec:pref}

There are  several parameters that control  the fundamental operations
of  the  GUI  and  the  text  displays.   These  can  be  edited  with
\inline{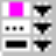}  and  \inline{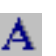},  respectively.
In \Tab{tab:prefs}, I describe the meaning of most of these keywords.

\begin{deluxetable}{ll}
  \tablecaption{Preferences\label{tab:prefs}}
  \tablehead{\colhead{Keyword} & \colhead{Description}}
  \startdata
  \multicolumn{2}{c}{GUI Preferences \cincludegraphics[height=11pt]{propsheet.pdf}}\\
  Eraser Size & eraser tool size (percent of image size)\\
  Painter Size & paint tool size (percent of image size)\\
  Zoom Speed & speed of the zoom for wheel motion\\
  Omit Zero & flag to omit zero in range calculation\\
  Min/max Method & algorithm for computing min/max\\
  RA/Dec Units & units of the RA/Dec display\\
  Scaling & grayscale distribution for direct image\\
  Invert Colors & flag to invert the direct image\\
  Range Color & color of the min/max box \cincludegraphics[height=11pt]{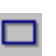}\\
  Range Opacity & opacity of the min/max box\\
  Range Linestyle & linestyle of border of the min/max box\\
  RGB Number & number of cached RGB triplets \tablenotemark{$\dagger$}\\
  Powscl Base & base of the power-law scaling\\
  \multicolumn{2}{c}{Text Preferences\tablenotemark{*} \cincludegraphics[height=11pt]{text.pdf}}\\
  Show & flag to show the text on the screen\\
  Color & RGB triplet of the text\\
  Fill background & flag to display a color behind the text\\
  Fill color & color of the background  
  \enddata  
  \tablenotetext{\dagger}{Should be larger than the maximum segmentation value.}
  \tablenotetext{*}{Other properties are ignored.}
\end{deluxetable}

\begin{figure}
  \begin{center}
    \includegraphics[width=2.25in]{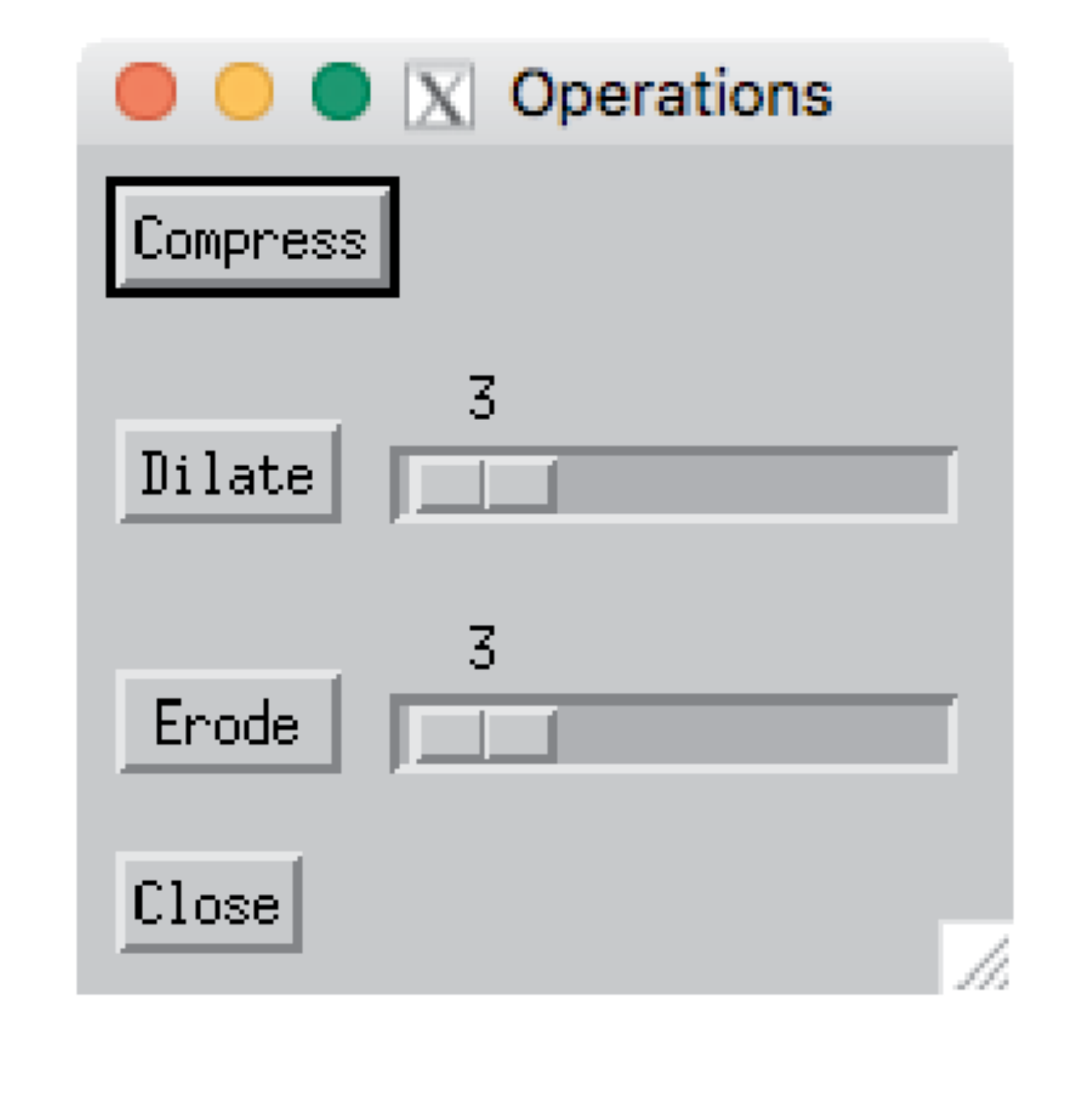}
    \caption{Operations tab.  The sub-GUI for various image processing
      options.    The  {\it   Compress}  button   will  compress   the
      segmentation map to  the lowest set of  positive integers, still
      reserving zero for the sky pixels.  The {\it Dilate} button will
      apply the  dilation operation  with size  given by  the adjacent
      slider.  The {\it Erode} button will apply the erosion operation
      with  the size  given by  the adjacent  slider.  Please  see the
      \texttt{IDL}  manual  for  more  details  on  these  latter  two
      operations.\label{fig:operations}}
  \end{center}
\end{figure}

\section{Additional Remarks and Functionality}\label{sec:add}

I have identified  a few places where the  current implementation does
not facilitate  certain actions that a  User may like.  I  may address
these in future versions, but am willing to work with Users wishing to
modify the codebase to enact these or other improvements.

\begin{description}
\item[Fixed window size] The size of the main graphics window is fixed
  to facilitate performance rendering and manipulating the RGB$\alpha$
  displays, which necessitated fixing the text sizes and margins.
  Either of these maybe relaxed in future, but Users wishing to force
  a different window size can easily adjust the \texttt{self.winsize}
  in the \texttt{init} method of the file
  \texttt{segeditor\_\_define.pro}.  This variable is in units of
  screen pixels.
\item[Segmentation map must exist] I have assumed that the
  segmentation map exists and will be modified. However, one may
  prefer to start with a blank segmentation map and simply draw the
  entire collection of regions.  If such functionality is required,
  then a simple workaround is to create a blank segmentation map {\it
    before} working with \texttt{SegEditor} and ensure that the
  world-coordinate system variables match those of the direct image.
  Of course, such a created image should likely have the
  \texttt{BITPIX} keyword set to an integer-like value.
\item[More interaction  in sub-GUIs]  There are several  sub-GUIs that
  can  be  spawned  from   within  \texttt{SegEditor},  however  their
  relative  interaction  with the  primary  graphics  window is  quite
  limited.  One  such example  of these limitation  is related  to the
  tabulation GUI, where it might be desirable to have the table update
  as regions are modified in the primary graphics window.  Conversely,
  it is often  advantageous to interact with segmentation  map via the
  table  (such as  deleting regions  in the  table view).   Until such
  changes are  implemented, it  is necessary  to refresh  the sub-GUIs
  and/or the primary graphics window as changes are made.
\end{description}

\section{Coda}\label{sec:coda}

I hope others find this software useful, and if so, then I would
gratefully appreciate a reference to this report.  I encourage any
Users to report bugs and/or suggestions.   I distribute the software
according to The MIT License.

%\begin{center}
%\begin{verbatim}
%The MIT License (MIT)
%
%Copyright (c) 2018 Russell Ryan
%
%Permission is hereby granted, free of 
%charge, to any person obtaining a copy 
%of this software and associated documentation 
%files (the "Software"), to deal in the 
%Software without restriction, including 
%without limitation the rights to use, copy, 
%modify, merge, publish, distribute, 
%sublicense, and/or sell copies of the 
%Software, and to permit persons to whom the 
%Software is furnished to do so, subject to 
%the following conditions:
%
%The above copyright notice and this 
%permission notice shall be included in all 
%copies or substantial portions of the 
%Software.
%
%THE SOFTWARE IS PROVIDED "AS IS", WITHOUT 
%WARRANTY OF ANY KIND, EXPRESS OR IMPLIED,
%INCLUDING BUT NOT LIMITED TO THE WARRANTIES 
%OF MERCHANTABILITY, FITNESS FOR A PARTICULAR 
%PURPOSE AND NONINFRINGEMENT. IN NO EVENT 
%SHALL THE AUTHORS OR COPYRIGHT HOLDERS BE 
%LIABLE FOR ANY CLAIM, DAMAGES OR OTHER 
%LIABILITY, WHETHER IN AN ACTION OF CONTRACT, 
%TORT OR OTHERWISE, ARISING FROM, OUT OF OR 
%IN CONNECTION WITH THE SOFTWARE OR THE USE 
%OR OTHER DEALINGS IN THE SOFTWARE.
%
%\end{verbatim}
%\end{center}

%This software is provided as is without any 
%warranty whatsoever.  Permission to use, copy, 
%modify, and distribute modified or unmodified 
%copies is granted, provided this copyright and 
%disclaimer are included unchanged.

\acknowledgments

I am very grateful to Duho Kim, Teresa Ashcraft, Andrea Bellini, Seth
Cohen, and Joe Hunkeler for their help in debugging and useful
suggestions.

\software{IDL}

\hfill 

\newpage

\begin{deluxetable*}{cll}
  \tablecaption{Button Controls\label{tab:buttons}}
  \tablehead{\colhead{} & \colhead{Name} & \colhead{Description}}
  \startdata
  \multicolumn{3}{c}{File Controls}\\
  \cincludegraphics[width=12pt]{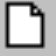} & New & delete all segmentation regions\\
  \cincludegraphics[width=12pt]{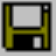} & Save & save the current segmentation map\\
  \multicolumn{3}{c}{Preferences}\\
  \cincludegraphics[width=12pt]{propsheet.pdf} & Image Prefs & edit the image properties\\
  \cincludegraphics[width=12pt]{text.pdf} & Text Prefs & edit the properites of the text displays\\
  \multicolumn{3}{c}{Undo/Redo}\\
  \cincludegraphics[width=12pt]{stepback.pdf} & Undo & undo the last region manipulation\\
  \cincludegraphics[width=12pt]{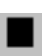} & Clear History & delete the cache of undo/redo commands\\
  \cincludegraphics[width=12pt]{step.pdf} & Redo & redo the last region manipulation\\
  \multicolumn{3}{c}{Left Mouse Button Control}\\
  \cincludegraphics[width=12pt]{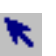} & Default & basic cursor state for pan, zoom, and pixel values\\
  \cincludegraphics[width=12pt]{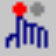} & Delete & left click to delete a region\\
  \cincludegraphics[width=12pt]{group.pdf} & Merge & left click and drag to merge two regions\\
  \cincludegraphics[width=12pt]{button.pdf} & Erase & left click and drag to erase segmentation regions under the cursor\\
  \cincludegraphics[width=12pt]{segpoly.pdf} & Draw & left click and drag to draw a new segmentation region\\
  \cincludegraphics[width=12pt]{drawing.pdf} & Unmerge & left click to unmerge a region from a parent\\
  \cincludegraphics[width=12pt]{paint.pdf} & Paint & left click and drag to paint the current segmentation region on the image\\
  \cincludegraphics[width=12pt]{rectangl.pdf} & Range & left click and drag to draw a box and compute min/max range from pixels inside the box\\
  \multicolumn{3}{c}{Right Mouse Button}\\
  \cincludegraphics[width=12pt]{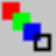} & Scale & toggle controls  right mouse button (engaged: opacity between images, disengaged: \texttt{ds9}-like scaling)\\
  \multicolumn{3}{c}{Miscellaneous Controls}\\
  \cincludegraphics[width=12pt]{dm.pdf} & Tabulate & view the region properties in a tabular format\\
  \cincludegraphics[width=12pt]{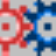} & Operations & open a sub-GUI for basic image operations (see \Fig{fig:operations})\\
  \cincludegraphics[width=12pt]{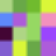} & Random RGBs & set a unique set of random RGB triplets for the segmentation map\\
  \cincludegraphics[width=12pt]{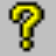} & Help & open a basic help display\\
  \enddata
\end{deluxetable*}

\end{document}